# Structure of Carbon Nanotube-dendrimer composite

V.Vasumathi[1], Debabrata Pramanik[1], A. K. Sood[2] and Prabal K Maiti[1*]

[1]Centre for Condensed Matter Theory, Department of Physics, Indian Institute of Science, Bangalore-560012, India.

[2]Department of Physics, Indian Institute of Science, Bangalore 560012, India

*Abstract*

Using all atomistic molecular dynamics (MD) simulations we report the microscopic picture of the nanotube-dendrimer complex for PAMAM dendrimer of generation 2 to 4 and carbon nanotube of chirality (6,5). We find compact wrapping conformations of dendrimer onto the nanotube surface for all the three generations of PAMAM dendrimer. The degree of wrapping is more for non-protonated dendrimer compared to the protonated dendrimer. For comparison we also study the interaction of another dendrimer, poly(propyl ether imine) (PETIM), with nanotube and show that PAMAM dendrimer interacts strongly as compared to PETIM dendrimer as is evident from the distance of closest approach as well as the number of close contacts between the nanotube and dendrimer. We also calculate the binding energy between the nanotube and the dendrimer using MM/PBSA methods and attribute the strong binding to the charge transfer between them. Dendrimer wrapping on CNT will make it soluble and can act as an efficient dispersing agent for nanotube.

[*] maiti@physics.iisc.ernet.in

*1. Introduction*

Carbon nanotubes (CNTs) constitute a class of nanostructures which possess exceptional mechanical[1, 2], electrical[1, 3] and optical properties[1-3]. CNTs are promising materials for polymer composites[4-6], energy conversion[7-9], sensing[10-12] and biological application[13-17]. However, processing CNT´s and introducing them into real applications is severely limited by their low solubilities in organic solvents. In recent past, two general strategies have been explored for improving the CNT solubility: (i) covalent functionalization of sp2 carbons at the sidewalls with organic molecules[18-24] and (ii) non-covalent functionalization through supramolecular interactions[25-32] (eg. π-π stacking interactions). Covalent functionalization of CNT through sidewalls has been shown to disturb the conjugated π-electrons of the tubes and in turn affect their properties. In comparison, functionalization of CNTs through non-covalent interactions helps retain their properties. In this context, functionalizing CNT through hyperbranched polymers like dendrimers is promising strategy to achieve better solubility of CNTs[33]. Dendrimers are very interesting macromolecules which possess a well-defined, three-dimensional structure and a multivalent exterior that can be modified with surface groups[34-36]. A fruitful association of these two wonderful nanoscale objects can give rise to many exciting material properties which can have wide range of applications[27, 37-45]. The interaction of CNT with dendrimers is not fully understood and only one study addresses the complexation between poly(propyl ether imine) (PETIM) dendrimer and the nanotubes[46]. However no microscopic study exits which deals with the noncovalent interaction between the more well-known PAMAM dendrimer and CNTs. Apart from helping better solubility, nanotube dendrimer composite can be used for DNA and siRNA delivery with better efficiency. Pristine nanotube has also been shown to be toxic compared to chemically functionalized CNTs[47-51]. There are speculation that dendrimer wrapped nanotube can be better suited for the delivery of nucleic acid with less toxicity[41, 52].

With this goal in mind, we have simulated the microscopic structure and interaction of the CNTs with protonated and non-protonated PAMAM dendrimer of generations 2 to 4 (G2, G3 and G4). For comparison purpose, we also present results for nanotube-PETIM dendrimer interactions of generation 3 and 4. Our all atom molecular modeling studies give a quantitative estimate of the binding affinity of different types of dendrimer (protonated and non-protonated) with the single walled carbon nanotube and also provide microscopic understanding as to what happens in going from one type of dendrimer architecture to another type. The rest of the paper is organized as follows: in the next section we give the details of system preparation as well as simulation conditions. We discuss various results coming out from our all atom MD simulations in section 3. Finally in section 4 we give a summary of the results and conclusions.

## 2. Simulation Details:

Initial structures of the protonated (corresponding to neutral pH) and non-protonated (corresponding to high pH) PAMAM and PETIM dendrimers of various generations were taken from our previous studies[53-58]. We have used the PMEMD software package[59] with the all atom AMBER03 force filed (FF)[60] for nanotube and Dreiding FF[61] for the dendrimers. Using the LEAP module in AMBER the protonated and non-protonated dendrimers of a given generation were placed in the vicinity of a nanotube of chirality (6,5). The length of the nanotube is 36 Å and the diameter is 7.5 Å. Appropriate numbers of Cl- counter ions were added to neutralize the positive charge on the terminal amines of the protonated dendrimers. The resulting structures were then solvated in TIP3P water box, extending 15 Å from the nanotube in all the three directions. The resulted nanotube plus dendrimer complex in a solvated form contains the nanotube, dendrimers of a given generation, water and Cl- counter ions in the case of a protonated dendrimer. Thus, we have generated nanotube- PAMAM or PETIM dendrimer complex using both protonated and non-protonated dendrimer of generations 2, 3 and 4 systems. Further, we compare our nanotube-PETIM dendrimer complex results with available experimental results on nanotube-PETIM dendrimer. In order to compare the results of nanotube-PAMAM dendrimer interaction with nanotube- PETIM dendrimer, one should be careful with generation labeling as the labeling of the generation for PAMAM and PETIM dendrimer is different. For PAMAM the labeling is as follows: core, 0, 1, 2, 3, .etc whereas for PETIM the labeling is: core, 1, 2, 3, etc. For comparison we should consider the generation which has the same number of terminal amines and hence n-th generation of PAMAM will be compared with the (n+1)th generation of PETIM. Hence, the results of nanotube-PAMAM dendrimer of nth generation are compared with the results of nanotube-PETIM dendrimer of (n+1)th generation. The details of the simulation conditions are given in table 1.

Each resulting solvated structures were subjected to 1000 steps of steepest descent minimization of the potential energy, followed by 2000 steps of conjugate gradient minimization. During this minimization, the nanotube and dendrimers were fixed in their starting conformations using harmonic constraints with a force constant of 500 kcal/mol/ $Å^2$. Due to the constraint on nanotube-dendrimer complexes, the water molecules reorganized themselves to eliminate bad contacts with the nanotube and dendrimers. The minimized structures were then subjected to 40 ps of MD, using a 2 fs time step for integration. While doing the MD, the system was gradually heated from 0 to 300 K using weak 20 kcal/mol/ $Å^2$ harmonic constraints on the solute to its starting structure. This allows for slow relaxation of the built nanotube-dendrimer structures. In addition SHAKE constraints[62] using a geometrical tolerance of 5 x$10^{-4}$ Å were imposed on all covalent bonds involving hydrogen atoms. This is needed to prevent changes in the NH and OH bonds from disrupting associated hydrogen bonds. Subsequently, MD was performed under constant pressure - constant temperature conditions (NPT), with temperature regulation achieved using the Berendsen weak coupling method[63] (0.5 ps time constant for heat bath coupling and 0.2 ps pressure relaxation time). This was followed by another 5000 steps of conjugate gradient minimization while decreasing the force constant of the harmonic restraints from 20 kcal/mol/ $Å^2$ to zero in steps of 5 kcal/mol/$Å^2$. The simulation protocol produce very stable MD trajectory for such complex system and earlier was found to be adequate to study complexation of ssDNA with dendrimer as well as large DNA and siRNA nanostructure in solution[64-66]. Finally, for analysis of structures and properties, we carried out 20-30 ns of NVT MD using a heat bath coupling time constant of 1 ps.

*3. Results and Discussion*

To characterize the structure of nanotube-dendrimer complex we have calculated the following quantities.

1. Center of mass distance between the nanotube and dendrimer

2. Number of contacts between dendrimer and the nanotube

3. Radial distribution function between the carbon atom of nanotube and terminal amines of dendrimers

4. van der Walls energy of interaction between the nanotube and dendrimers

5. Charge transfer between dendrimer and the nanotube

*Center of mass distance:*

All atoms MD simulations reveal that, dendrimers takes several nanoseconds to form a compact wrapping conformation onto the nanotube surface. Strong van der Waals interaction between the nanotube and the dendrimer is the driving for such complexation in contrast to electrostatic interaction which drives the complexation between dendrimer and nucleic acids [66, 67]. Figure 1 shows the center of mass (COM) distance between the carbon nanotube and the dendrimer. The COM distance decreases with simulation time. The distance of closest approach for PAMAM dendrimer as reflected by the COM distance is very similar for both the non-protonated and protonated cases. Also the COM distance increases with the increase in the dendrimer generation. Compared to PETIM dendrimer, PAMAM dendrimer comes closer to the dendrimer surface as is reflected by the smaller COM distance. For example, for G2 PAMAM dendrimer the equilibrium COM distance between the dendrimer and CNT is 3 Å compared to 5 Å distance in the case of G3 PETIM dendrimer and CNT. This could be due to the larger branch length (see fig 2) in case PETIM dendrimer compared to PAMAM dendrimer.

Figure 3 shows instantaneous snapshots of nanotube-dendrimer complex for generation 2(G2) non-protonated PAMAM dendrimers and CNT in a few ns interval to show the various stages of the dendrimers binding process on the surface of the nanotube. We observed similar behavior for all the other cases and the instantaneous snapshots for the case G3 PAMAM-CNT and G4 PAMAM-CNT complex are shown in figures 4 and 5 respectively. Our simulation results show that wrapping is more for the case of non-protonated dendrimers of all generations for both types of dendrimers, suggesting that interaction of dendrimers with nanotube is maximum for non-protonated dendrimers than the protonated dendrimers. To get a molecular origin for this binding patterns we have calculated the number of contacts between the dendrimer and CNT in the next section.

*Number of close contacts between nanotube and dendrimer:*

We have calculated the number of close contacts between nanotube and dendrimer, by using[66],

$$N_c(t) = \sum_{i=1}^{N_{CNT}} \sum_{j=1}^{N_{Den}} \int_{r_i}^{r_i+3\text{Å}} \delta(r(t) - r_j(t)) dr$$

Here, $N_{CNT}$ and $N_{Den}$ are the total number of nanotube and dendrimer atoms respectively, and $r_j$ is the distance of $j^{th}$ atom of dendrimer from $i^{th}$ atom of the nanotube. The delta function ensures counting of all dendrimer atoms within 3 Å from nanotube atoms. Figures 6 (a) and (b) show the number of close contacts ($N_c$) between the nanotube and PAMAM/PETIM dendrimer for non-protonated and protonated case. The number of close contacts increases as generation number increases for both type of dendrimers. Further we observe that the number of close contacts is slightly higher for nanotube-PAMAM dendrimer complexes when compared to the number of close contacts with PETIM dendrimer. Also non-protonated dendrimer has larger number of contacts compared to protonated case for both PAMAM and PETIM dendrimer. This is consistent with our earlier observation that COM distance for the non-protonated dendrimer is smaller compared to the protonated case.

*Radial Distribution Function:*

What is the molecular origin of the fact that PAMAM dendrimers have better binding affinity to nanotube compared to PETIM dendrimer? To gain a deeper understanding of this binding affinity for two different dendrimer architectures as a function of the dendrimer generations we have calculated the radial distribution function between the carbon atoms of the nanotube and the terminal amines of the dendrimers using the configurations of the last 6 ns long simulation. The RDF profiles of terminal amines of dendrimer with the carbon atoms of the nanotube for all the complexes are shown in Figure 7. Main observations are as follows: (i) the RDF is high for the non-protonated dendrimer for any given generation for both PAMAM and PETIM dendrimers, (ii) the distribution increases with the increase in the dendrimer generation for protonated and non-protonated dendrimers of both types. Another surprising observation is that RDF profile for non-protonated PAMAM dendrimer of generation G3 and G4 have approximately same peak height for the first peak. This might indicate that for the length of the nanotube used in our simulation (36 Å) G3 non-protonated dendrimer can give enough coverage to wrap the whole nanotube. Comparing RDF profiles for PETIM and PAMAM dendrimer we find that the distributions of terminal amines are high for PAMAM dendrimers compared to the PETIM dendrimers. This may be due to the variations of the branch length of the dendrimers. The branch length of the PETIM dendrimer is larger (9 Å) when compared to PAMAM dendrimer (7 Å) (see Fig 2), making PAMAM dendrimer more compact (see Fig.8).

*Binding Energy:*

So far we have analyzed various structure based indicators to quantify the binding of dendrimer to CNTs. To understand the thermodynamics of this binding we have calculated the binding energy for the various nanotube-dendrimer complexes studied in this work using Molecular Mechanics/Poisson-Boltzmann Surface Area method (MM-PBSA) module of AMBER9. In general the binding free energy for the non-covalent association of two molecules may be written as $\Delta G(A+B \rightarrow AB) = G_{AB} - G_A - G_B$ .

For any species on the right hand $G(X) = H(X) - TS(X)$ side, the above binding energy can be written as $\Delta G_{bind} = \Delta H_{bind} - T\Delta S_{bind}$ .

where $\Delta H_{bind} = \Delta E_{gas} + \Delta G_{sol}$ . $\Delta H_{bind}$ is the change in enthalpy and is calculated by summing the gas-phase energies ($\Delta E_{gas}$) and solvation free energies ($\Delta G_{sol1}$). Here $E_{gas} = E_{ele} + E_{vdw} + E_{int}$ ; $E_{ele}$ is the Electrostatic energy calculated from the Coulomb potential, $E_{vdw}$ is the non-bond van der Walls energy and $E_{int}$ is the internal energy contribution from bonds, angles and torsions. $G_{sol} = G_{es} + G_{nes}$, where $G_{es}$ is the electrostatic energy calculated from a Poisson-Boltzmann (PB) method and $G_{nes}$ is the non-electrostatic energy calculated as where $\gamma \times SASA + \beta$ $\gamma$ is the surface tension parameter ($\gamma = 0.00542$ kcal/ Å$^2$) and $SASA$ is the solvent-accessible surface area of the molecule.

From the MMPB/SA calculation we observe that the change in the binding energy is due to van der waals interaction. For example, the energy components $E_{ele}$ and $E_{int}$ = 0 and $E_{vdw}$ =  in the case of PAMAM  protonated dendrimer of generation

As mentioned earlier CNT is uncharged and the dendrimer-CNT interaction is driven by the dispersive van der Waals energy. The main contribution for the enthalpy therefore comes from van der Waals energy. Table 2 shows the van der Waals energy for all the cases and we find that the van der Waals energy is increasing as the generation of the dendrimer is increased for both types of dendrimers. Also, binding energy is more for non-protonated case as compared with the protonated case for both the PAMAM and PETIM dendrimers. Further, we observed that, the energy is more for PMAM dendrimer when compared with the PETIM dendrimer.

*Charge transfer*:

To have more microscopic origin of nanotube interacts with PETIM and PAMAM dendrimers we have also calculated the charge transfer between the nanotube and dendrimer. The charge transfer was computed using the QeQ charge equilibration method[68] using few equilibrium configurations of the dendrimer-nanotube complex from our long MD simulations. The QeQ calculations were performed using Cerius2 from Accelrys. QeQ was originally proposed by Rappe and Goddard to calculate the atomic charges for molecular systems from the electrostatic energy which takes into account molecular geometry. The main ingredients in this method are atomic ionization potentials, electronegativity and atomic radii. For molecular systems the electrostatic energy is expressed as a sum of intra-atomic contributions and inter-atomic interaction between pair of atoms. Atomic charge distribution is obtained by minimizing the electrostatic energy. Table 3 gives the charge transfer from the dendrimer to the nanotube for all the systems studied in this work. We observe the remarkable changes in the charge-transfer for nanotube-PETIM dendrimer complex depending on the dendrimer generations. Further, the amount of charge transfer for the non-protonated case is almost twice compared to the protonated case (in particular for the case of G3 and G4 PETIM dendrimers). This is consistent with the previous experimental results[46]. But, the amount of charge transfer for PAMAM dendrimer-nanotube complex does not change much with the increase in dendrimer generation. Our results also reveal that charge transfer interaction is more for the nanotube-PETIM complex compared to the nanotube-PAMAM dendrimer complex. This result is in contrast to our other findings that show stronger interaction and binding affinity for nanotube-PAMAM complex compared to the nanotube-PETIM dendrimer complex. This variation could due to the difference in chemical nature of the branches of the two dendrimers. Figure 2 shows the branches of the PETIM and PAMAM dendrimers. From the figure one can say that hydrogen bond can form for PAMAM dendrimer such that free lone pair electrons present in oxygen may not transfer to nanotube. However, for PETIM dendrimer there is no possibility of creating hydrogen bond and hence more electrons from the lone pair present in oxygen atoms may transfer to the nanotube. Hence, we observe the remarkable change in the charge-transfer for nanotube-PETIM dendrimer interaction and not for nanotube-PAMAM dendrimer interaction. From this result, we may conclude that nanotube-PETIM dendrimer complex can be used as photoactive surfaces in photovoltaics[69, 70].

*Summary and Conclusion:*

To summarize, we have studied the microscopic picture of the CNT-PAMAM dendrimer and CNT-PETIM dendrimer complex for different generations at high (non-protonated case) and neutral pH (protonated case) using all atom MD simulations. There is a strong van der Waals interaction between the CNT and the dendrimer mediated by the charge transfer between them. Over nanosecond long time scale this van der Waals interaction helps dendrimer wrap around the CNT to form a tight complex. For a given length of CNT, larger the dendrimer generation better is the complex formation as revealed by the RDF between CNT and dendrimer, as well as number of close contacts. Calculated RDF profile and binding energies show that, the interaction is stronger for the nanotube-non-protonated dendrimer complexes as compared to nanotube-protonated dendrimer complexes for both the PAMAM and PETIM dendrimers. Between the nanotube-PMAM dendrimer complexes and nanotube-PETIM dendrimer complexes, the nanotube-PAMAM dendrimer complexes shows maximum interaction and the interaction strength increases as the dendrimer generation is increased from G2 to G3. However interaction strength remains unchanged changing the generation from G3 to G4. These imply that size of the G3 PAMAM dendrimer is enough to wrap the nanotube of the length 36 Å and diameter of 7.5 Å. Our charge-transfer results suggest that the nanotube is charged more when it interact with the PETIM dendrimer and the charge increases with higher generation. From these results one can conclude that, nanotube-PAMAM dendrimer complexes are more stable than the nanotube-PETIM dendrimer complexes but there is no significant change in the charge of the nanotube. Hence, the nanotube-PAMAM dendrimer complexes can be used as efficient carrier for drug delivery and the nanotube-PETIM dendrimer complexes can be used as efficient photoactive surfaces in photovoltaics.

**Acknowledgements**
We thank DBT, India for financial support.

**Tables**

Table 1: Details of the simulation condition of the nanotube-dendrimer complexes studied in this work.

| CNT with | No of CNT atoms | No of dendrimer atoms | No of water molecules | Total no of atoms | Cl- ions | Box dimension (Å$^3$) |
|---|---|---|---|---|---|---|
| PAMAM-G2-non-protonated | 364 | 516 | 16021 | 48943 | 0 | 77 x 74 x 87 |
| PAMAM-G2-protonated | 364 | 532 | 17886 | 54570 | 16 | 81 x 77x 89 |
| PAMAM-G3-non-protonated | 364 | 1092 | 27626 | 83758 | 0 | 94 x 92 x 99 |
| PAMAM-G3-protonated | 364 | 1124 | 29441 | 89843 | 32 | 93 x 96 x 102 |
| PAMAM-G4-non-protonated | 364 | 2244 | 40932 | 125404 | 0 | 107 x 113 x 106 |
| PAMAM-G4-protonated | 364 | 2308 | 46840 | 143256 | 64 | 119 x 105 x 115 |
| PETIM-G3-non-protonated | 364 | 613 | 16634 | 50879 | 0 | 72 x 80 x 90 |
| PETIM-G3-protonated | 364 | 629 | 17221 | 52672 | 16 | 71 x 83x 90 |
| PETIM-G4-non-protonated | 364 | 1285 | 27964 | 85541 | 0 | 103 x 91 x 93 |
| PETIM-G4-protonated | 364 | 1317 | 28580 | 87453 | 32 | 104 x 90 x 95 |

Table 2: van der Waals energy determined by molecular dynamics along with MMPBSA of the nanotube-dendrimer complexes studied in this paper.

|  | PETIM (kcal/mol) | PAMAM (n-1)g (kcal/mol) |
|---|---|---|
| G3-protonated | -107.67 +/- 9.58 | -154.04 +/- 7.0 |
| G3-non protonated | -146.51+/- 5.76 | -181.84 +/- 6.0 |
| G4-protonated | -156.37 +/- 8.3 | -184.61 +/- 9.56 |
| G4-non protonated | -203.86 +/- 6.06 | -203.41 +/- 7.9 |
| G5-protonated |  | -241.94 +/- 9.1 |
| G5-non protonated |  | -287.24 +/- 7.6 |

Table 3: Charge transfer of the nanotube-dendrimer complexes of the all cases.

|  | PETIM | PAMAM (n-1)g |
|---|---|---|
| G3-protonated | -0.83 | 0.16 |
| G3-non protonated | -1.89 | -0.55 |
| G4-protonated | -1.35 | -0.09 |
| G4-non protonated | -2.11 | -0.6 |
| G5-protonated |  | -0.02 |
| G5-non protonated |  | -0.66 |

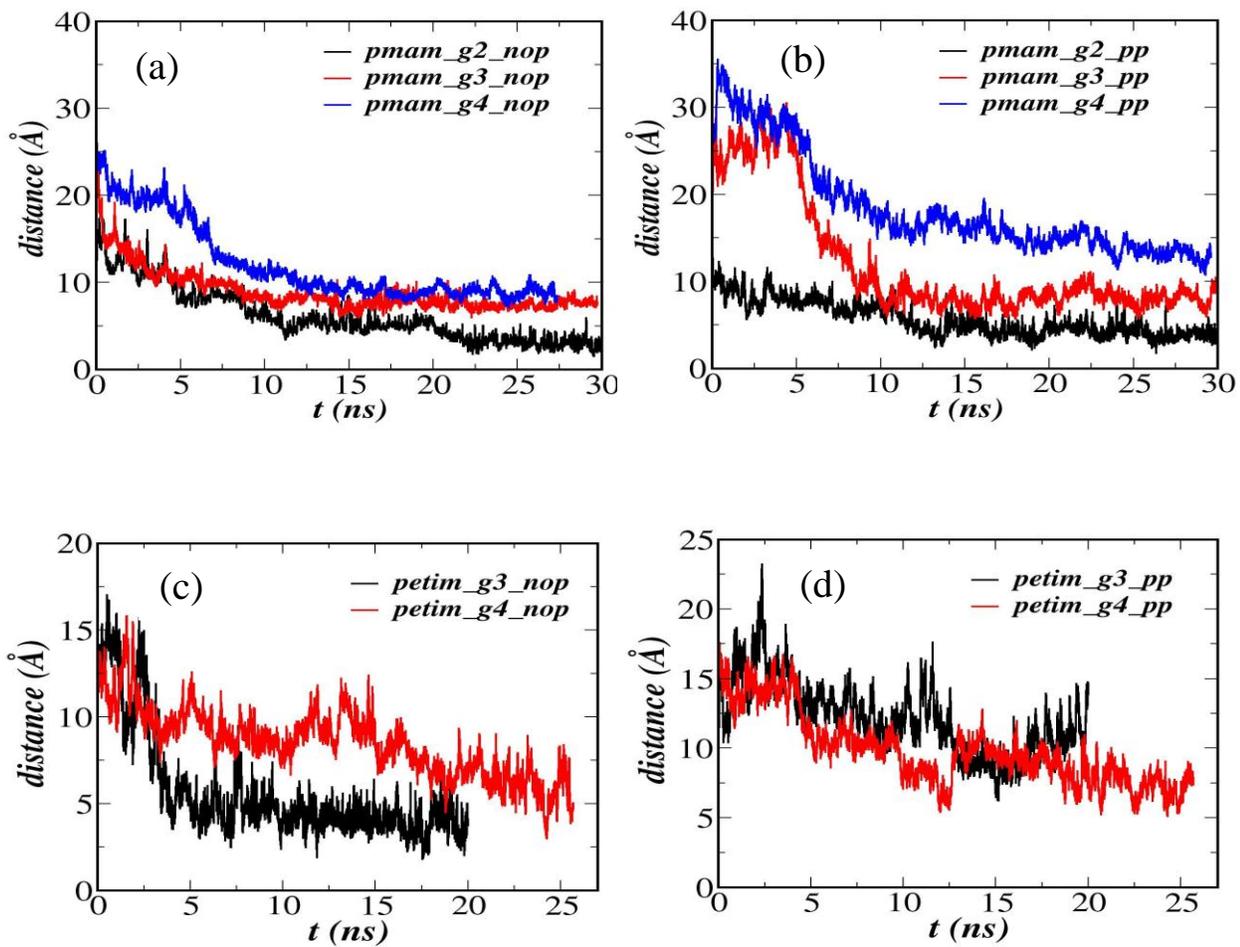

Figure 1: The time evolution of the distance between the center of mass (COM) of nanotube and the center of mass of dendrimers of (a) PAMAM non-protonated (b) PAMAM protonated (c) PETIM non-protonated and (c) PETIM protonated state.

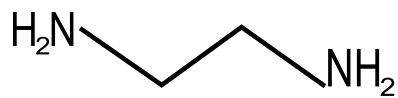

(a)

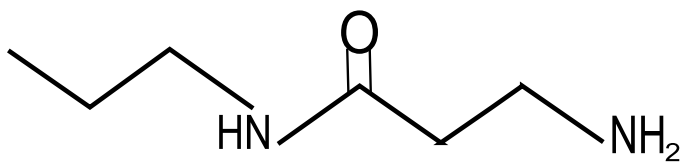

(b)

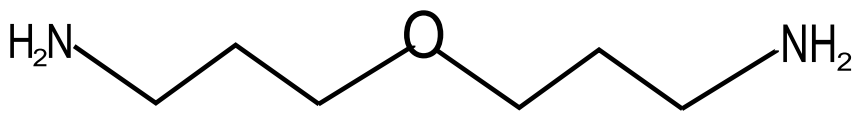

(c)

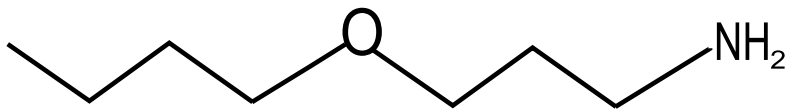

(d)

Figure 2: Schematic diagram of chemical structure of (a) core (4 Å) and (b) branch (7 Å) of PAMAM dendrimer (c) core (9 Å) and (d) branch (9 Å) of PETIM dendrimer.

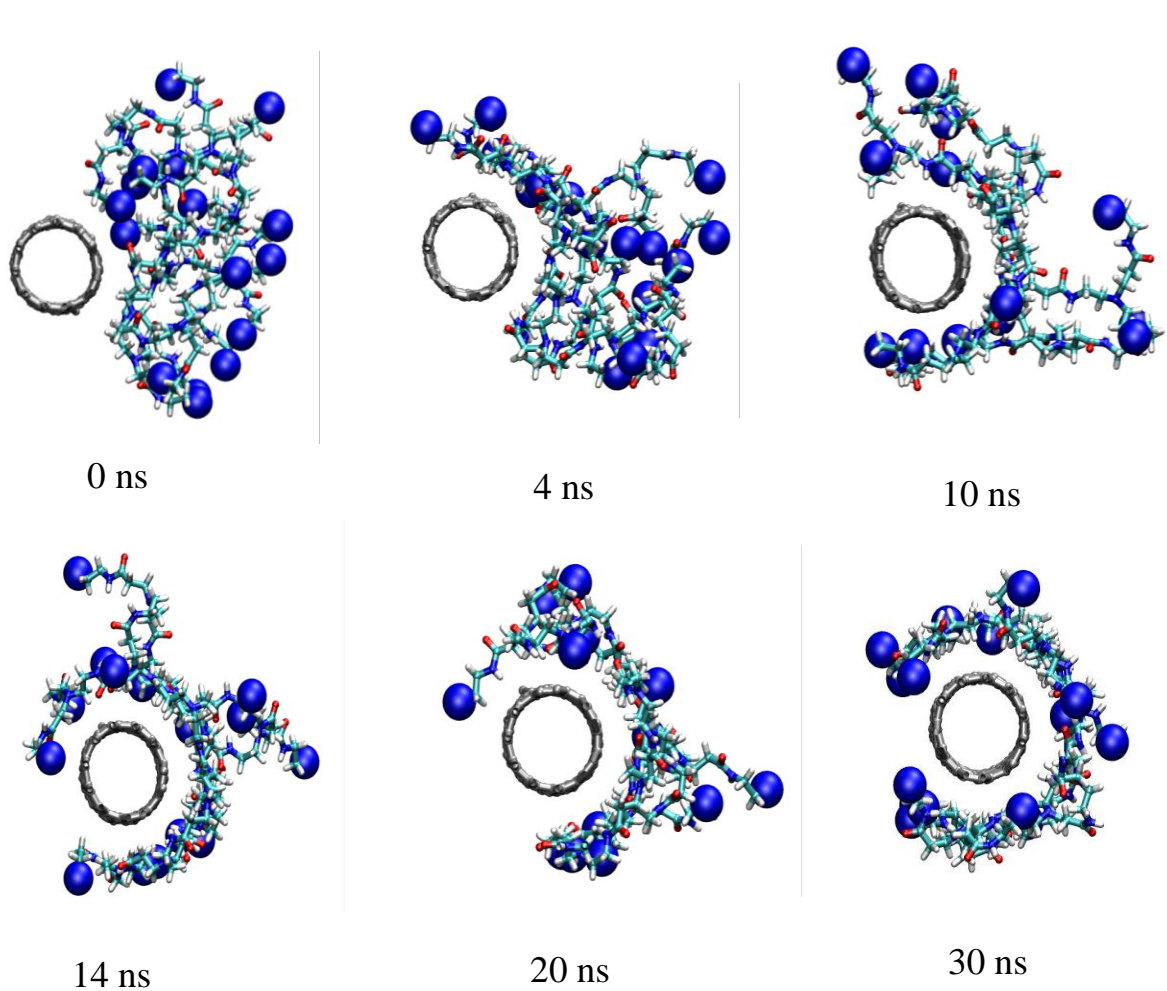

Figure 3: (a) Structure of nanotube-G2 PAMAM dendrimer complex for non-protonated case during various stages of the binding process at the interval of few ns. Blue spheres represent the terminal amines.

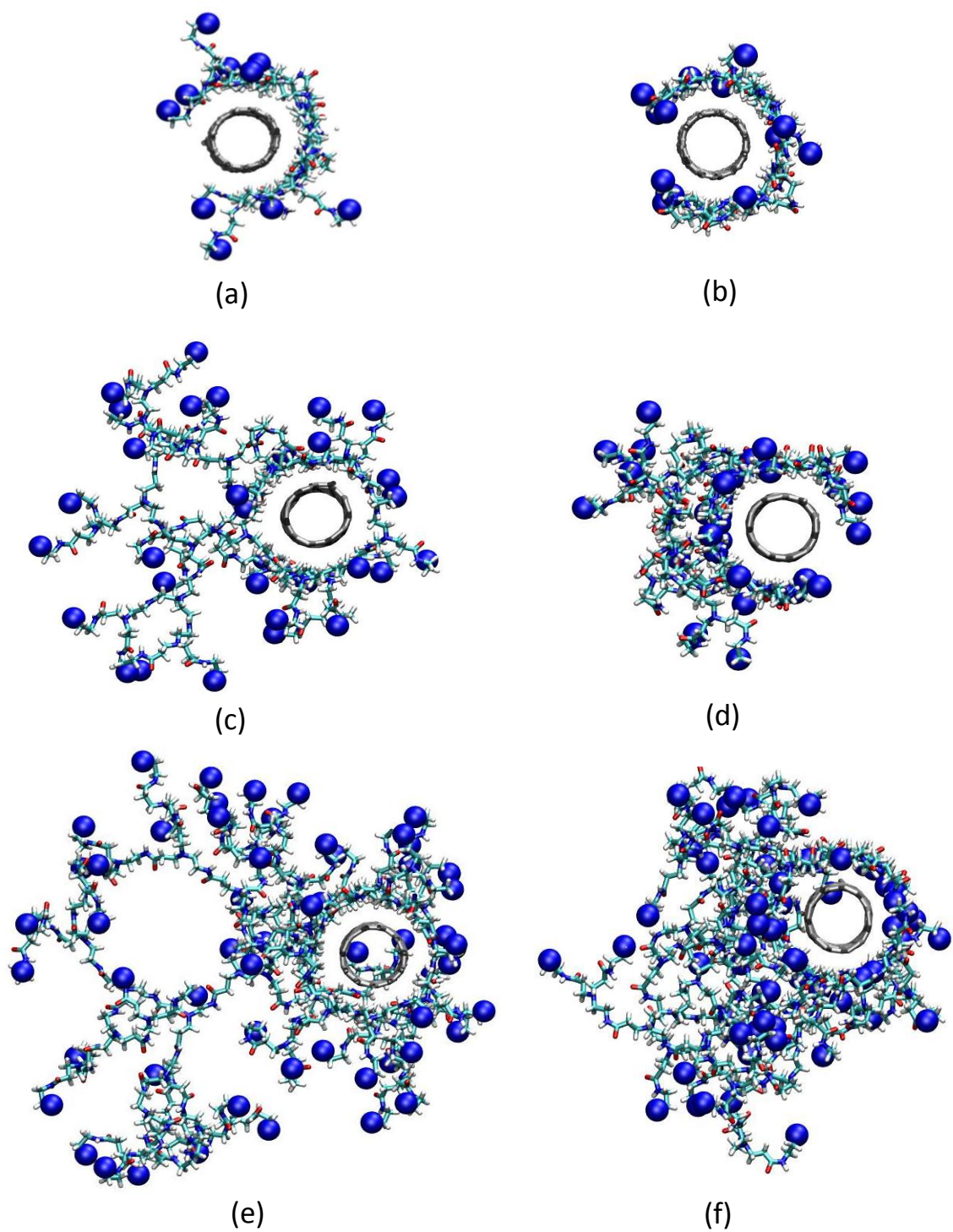

Figure 4: Equilibrated snapshot of the nanotube-PAMAM dendrimer complexes: nanotube with (a) protonated (b) non-protonated dendrimer of generation 2, (c) protonated (d) non-protonated dendrimer of generation 3, (e) protonated and (f) non-protonated dendrimer of generation 4.

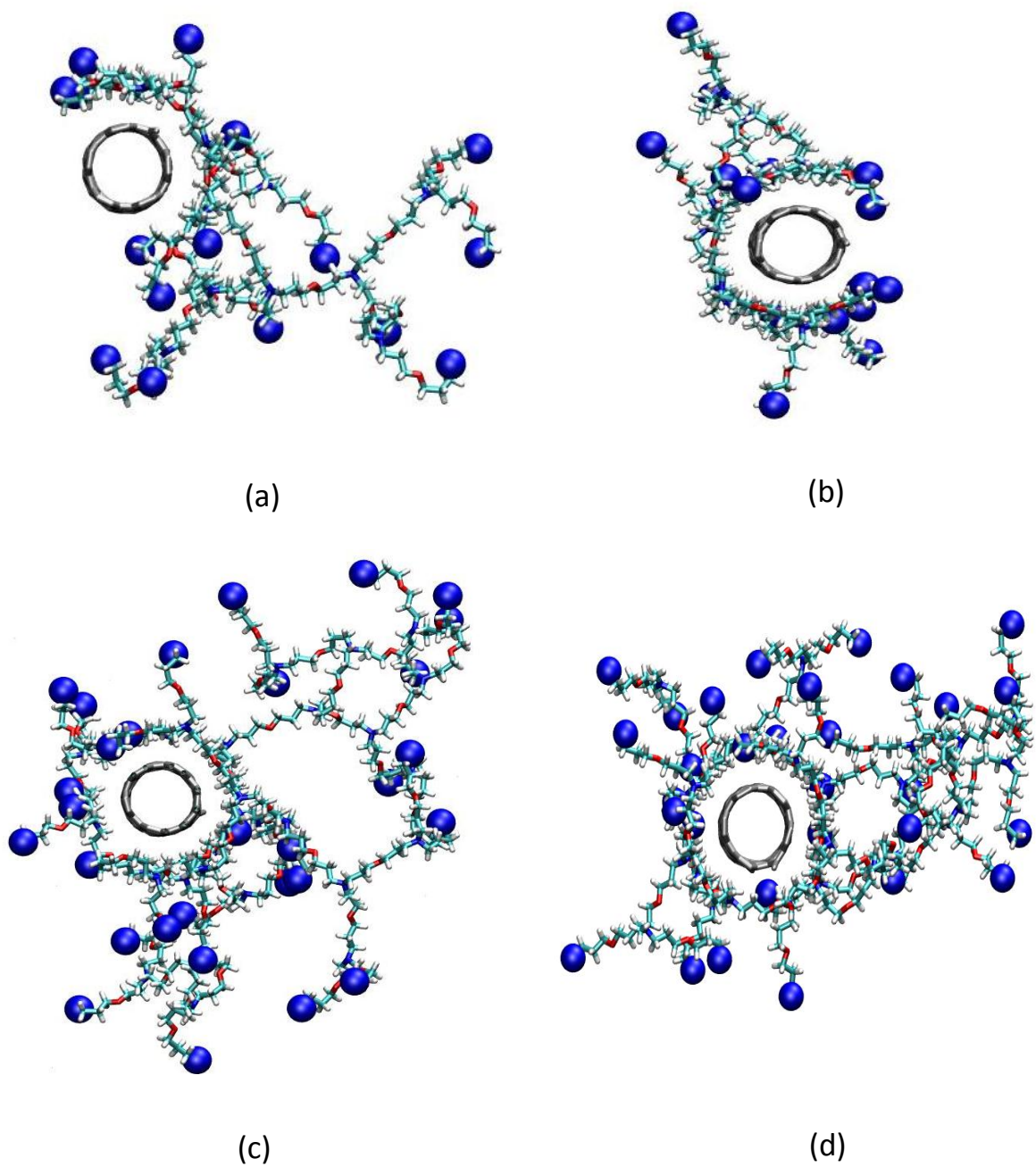

Figure 5: Instantaneous snapshots of the nanotube-PETIM dendrimer complexes: nanotube with (a) protonated (b) non-protonated dendrimer of generation 3, (c) protonated and (d) non-protonated dendrimer of generation 4.

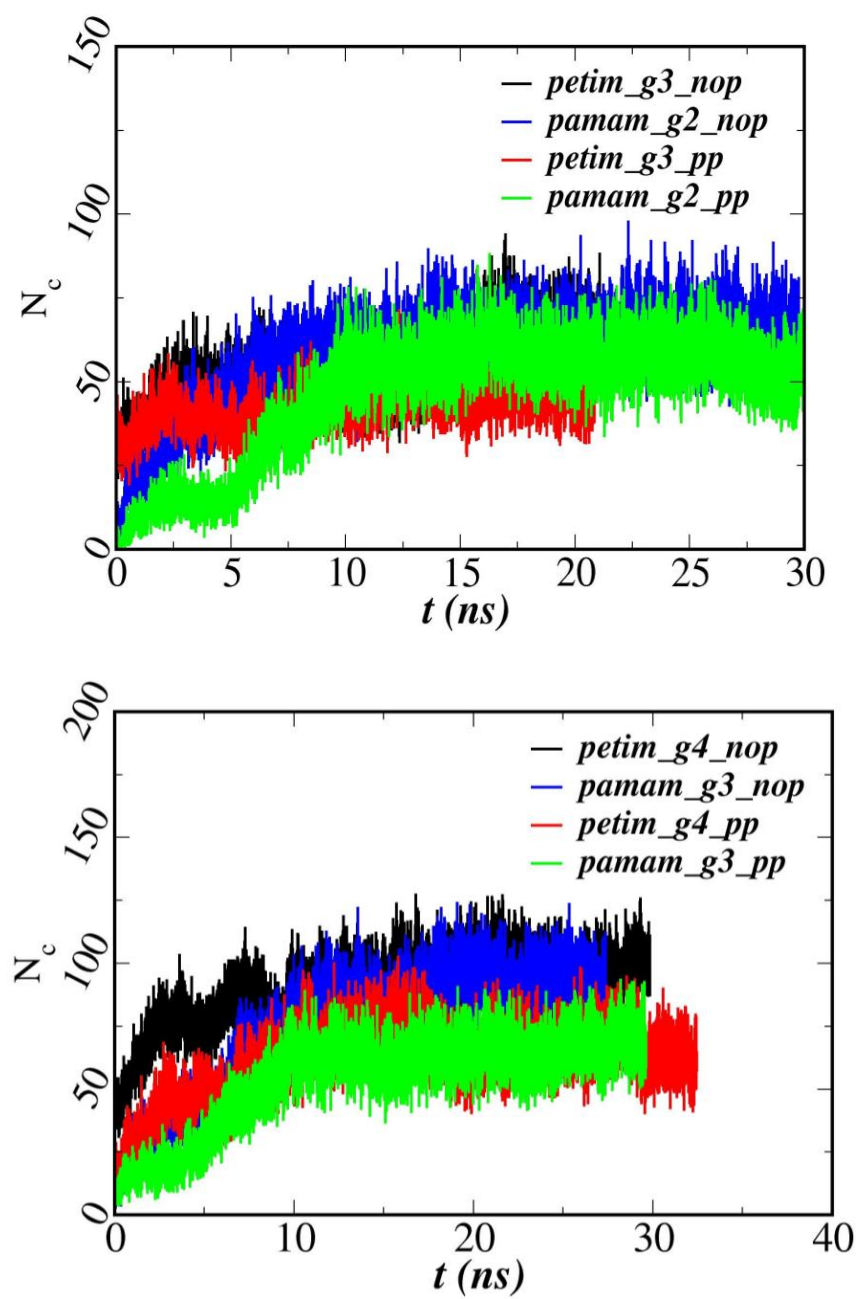

Figure 6: Variation of the number of close contact atoms between nanotube and dendrimer (any contact within 3 Å ) of (a) generation 2(PAMAM) and 3(PETIM) and (b) generation 3 (PAMAM) and 4 (PETIM).

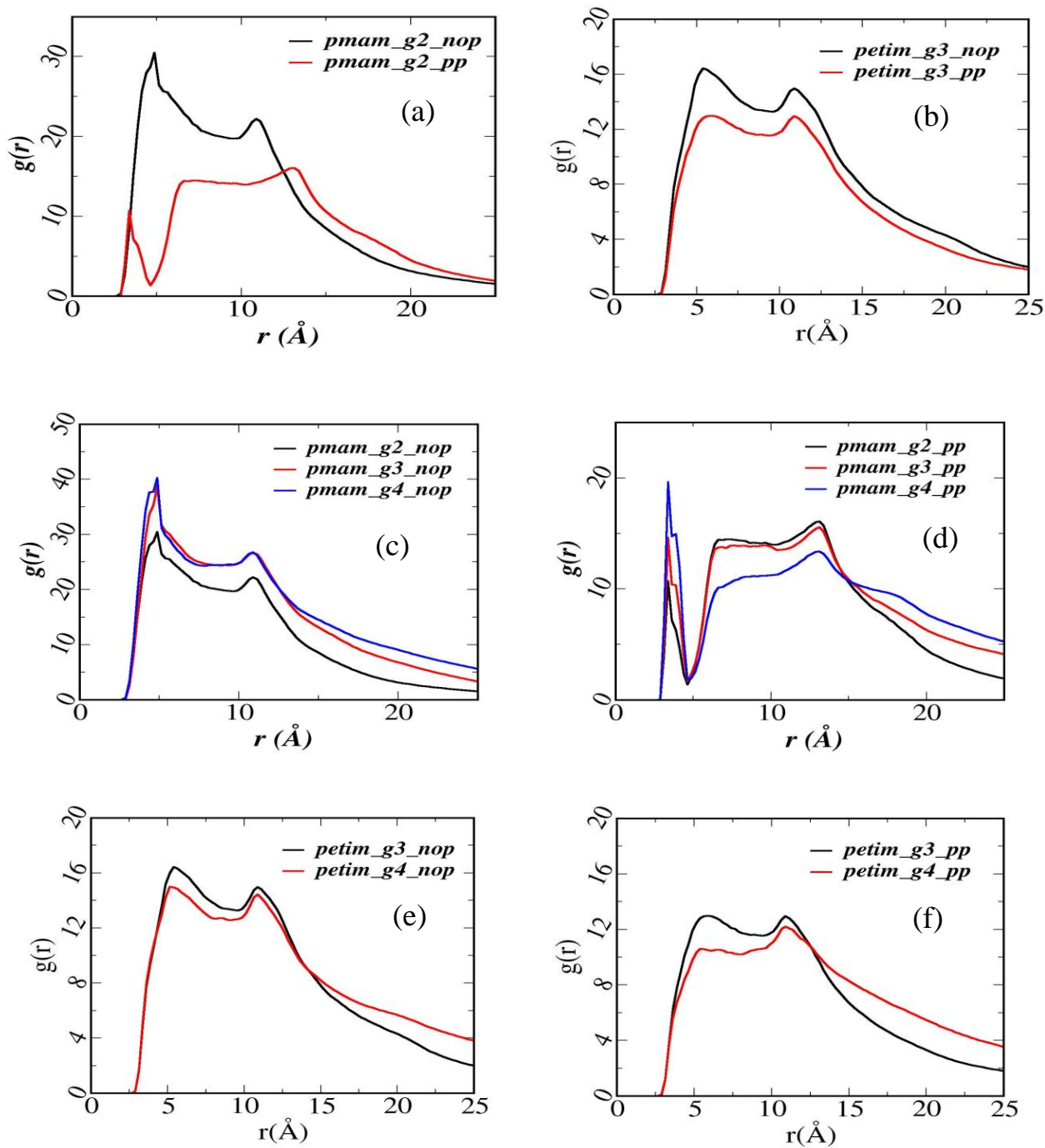

Figure 7: Radial distribution function (RDF) between carbons in nanotube and terminal amines in dendrimer of (a) G2 PAMAM (b) G3 PETIM (c) G2, G3 and G4 non-protonated PAMAM (d) generation 2,3,4 of protonated PMAM (e) G3 and G4 non-protonated PETIM (f) G3 and G4 protonated PETIM.

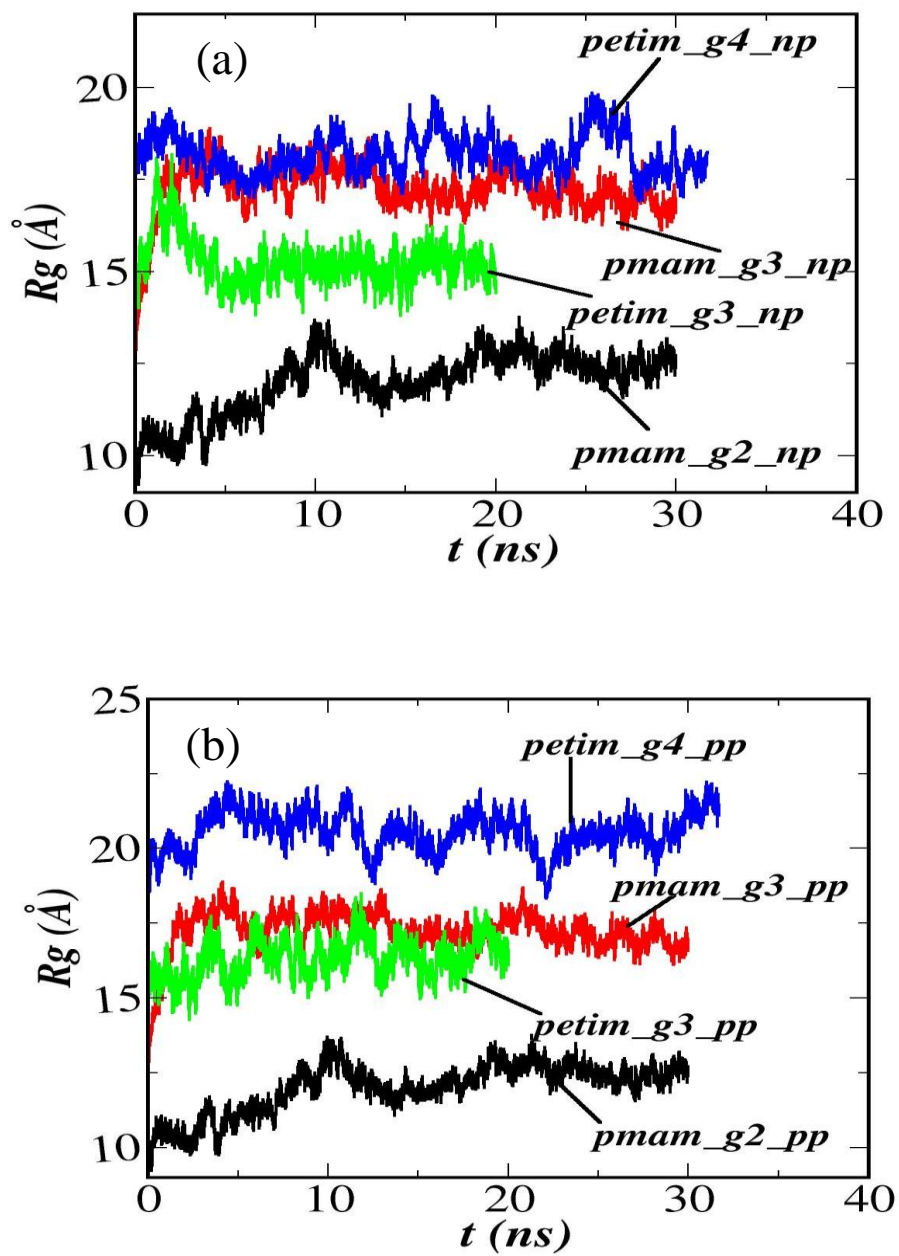

Figure 8: Evolution of the size of the PAMAM and PETIM (a) non-protonated dendrimer (b) protonated dendrimer showing the conformational change during the binding process.